\documentclass[aps,prl,twocolumn,groupedaddress]{revtex4-1}

\usepackage{graphicx}
\begin{document}


\title{Improving Dark Matter Axion Searches with Active Resonators}


\author{Gray Rybka}
\email[]{grybka@uw.edu}
\affiliation{University of Washington}


\date{\today}

\begin{abstract}
Axions are a well motivated candidate for dark matter.  The most sensitive experiments searching for dark matter axions rely on the coupling of axions to the electromagnetic resonances of a microwave cavity immersed in a strong magnetic field.  The sensitivity of the experiment is proportional to the $Q$ of the resonance that is coupled to axions.
To date, the resonators used in axion searches have all been passive, with $Q$s limited by power loss in the cavity walls.  I propose the use of active feedback resonators to increase the $Q$ of microwave cavity axion dark matter experiments by several orders of magnitude.
This should allow experiments to significantly increase the rate at which they can test potential axion masses and couplings.
\end{abstract}


\maketitle

The axion is a hypothetical particle that is both a candidate for dark matter and a result of the Peccei-Quinn solution to the strong CP problem~\cite{Peccei,Peccei_2,PhysRevLett.40.223,PhysRevLett.40.279,Preskill1983127,Abbott1983,ipser-sikivie}.
Axions with masses below $10^{-3}$ eV are particularly interesting because they could be produced in sufficient quantities to account for dark matter~\cite{Turner199067}.
For axions in this mass range the coupling between axions and photons, despite being exceptionally weak, provides the best chance of directly observing axion dark matter.

The most sensitive dark matter axion searches to date have been of the ``microwave cavity" type.
These experiments rely on the conversion of axions from the local dark matter halo into photons in a strong magnetic field.
This conversion is resonantly enhanced when the resonant frequency of the microwave cavity is equal to the frequency of the photons produced from the axion conversion~\cite{PhysRevLett.51.1415}.
Operation of these experiments involves slowly tuning the resonant frequency of the microwave cavity to explore different potential axion masses and searching for an excess of power deposited from dark matter axion conversion.
Microwave cavity experiments have been demonstrated to have the sensitivity required to detect optimistically coupled dark matter axions over a small mass range, but as of yet, axions have not been detected in a microwave cavity experiment~\cite{PhysRevLett.104.041301}.

The signal power in microwave cavity experiments is proportional to a number of factors, including the local density of dark matter, the strength of the magnetic field, the volume of the cavity, and the resonant quality factor $Q$ of the resonance being used.
The primary background is the thermal noise from the physical temperature of the cavity and the electronic noise from the first stage amplifier.  
The figure of merit for a microwave cavity experiment is the instantaneous axion signal to noise ratio (SNR).
Experiments with a larger SNR can be sensitive to more pessimistic axion photon couplings for a given frequency tuning speed or tune more quickly for a given axion photon coupling sensitivity.

Increasing the cavity $Q$ is one way to increase SNR in a microwave cavity experiment; the speed at which the cavity frequency can be tuned while remaining sensitive to a given axion photon coupling is linearly proportional to $Q$~\cite{Peng2000569}.
The presence of a strong magnetic field precludes the use of high-$Q$ superconducting cavities in these experiments, so the cavities are usually made of or coated with copper.
Loaded $Q$s of $10^5$ at 1 GHz have been achieved with copper cavities in axion experiments~\cite{Peng2000569}.  Higher frequency cavities typically have smaller $Q$s.

I present here a means of artificially increasing the $Q$ of the cavity resonance using active feedback in order to improve sensitivity to dark matter axions and increase the speed at which different axion masses can be tested.

\begin{figure}
\includegraphics[width=7cm]{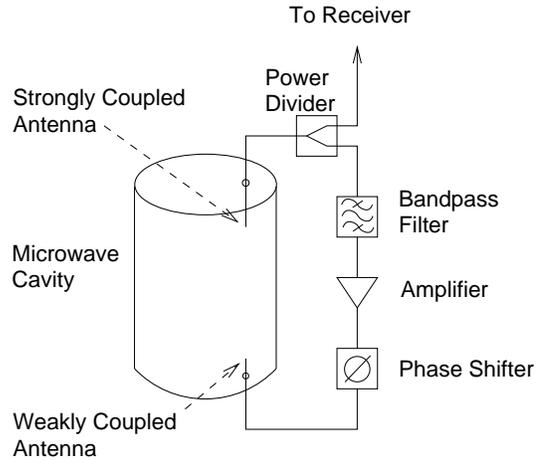}
\caption{\label{fig:experiment_schematic} Schematic of active feedback resonator for an axion experiment}
\end{figure}

\begin{figure}
\includegraphics[width=6cm]{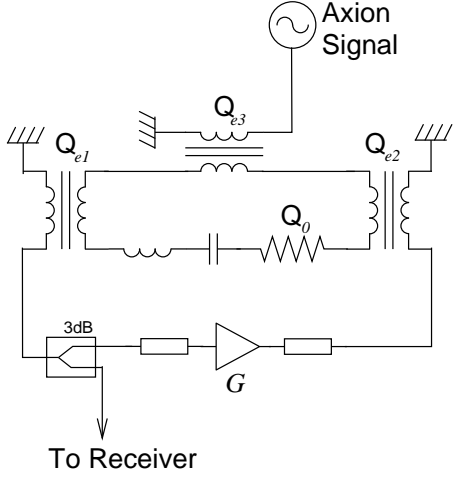}
\caption{\label{fig:equiv_circuit} Equivalent circuit to axion photon conversion in microwave cavity with active regeneration}
\end{figure}

\begin{figure}
\includegraphics[width=6cm]{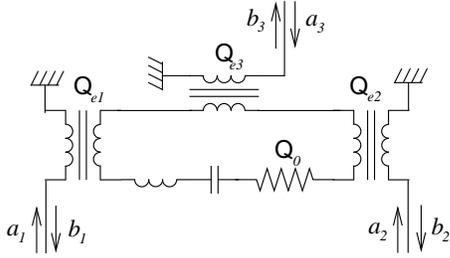}
\caption{\label{fig:amplitude_defs} Microwave cavity represented as three port device.}
\end{figure}

The use of active feedback resonators is a well established technique used in regenerative receivers~\cite{armstrong1914wireless}.
It involves connecting positive feedback to a resonator at the appropriate phase and amplification such that the power lost each oscillation to the output and damping effects is nearly completely replaced.  
This has the effect of increasing the $Q$ of the system.

A schematic of a microwave cavity axion experiment with active feedback is shown in Fig.~\ref{fig:experiment_schematic}.   Power from axion to photon conversion exits the cavity via a strongly coupled antenna, some of which is sent to a detector.  The remainder of the power is amplified, phase shifted, and then fed back into the cavity via a weakly coupled antenna.  A filter may also be needed to select the desired mode. 

To calculate the effect of the active feedback, I will use the equivalent circuit to the cavity-amplifier system shown in Fig.~\ref{fig:equiv_circuit}.  The cavity resonant mode is equivalent to an $LRC$ circuit with an unloaded $Q$ of $Q_0$ and the coupled antennas are equivalent to transformers with couplings $Q_{e1}$ and $Q_{e2}~$\cite{Montgomery:1948}.  $G$ is the combined gain of the amplifier and power divider.  There is also a phase shift between the two antennas around the loop, which will be denoted as $\delta$.  The coupling between the electromagnetic resonance in the cavity and dark matter axions is equivalent to a third antenna with coupling $Q_{e3}$.  The passive loaded $Q$ of the resonator is $Q_L=\left(Q_0^{-1}+Q_{e1}^{-1}+Q_{e2}^{-1}+Q_{e3}^{-1}\right)^{-1}$.

I can now treat the cavity as a three port device using amplitudes defined in Fig.~\ref{fig:amplitude_defs} with an $S$ matrix of
\begin{equation}
S_{\mathrm{cavity}}=
\left( \begin{array}{ccc}
		\frac{2Q_L}{Q_{e1}}-1 & \frac{2Q_L}{\sqrt{Q_{e1}Q_{e2}}} & \frac{2Q_L}{\sqrt{Q_{e1}Q_{e3}}}  \\
		\frac{2Q_L}{\sqrt{Q_{e1}Q_{e2}}} & 1-\frac{2Q_L}{Q_{e2}} & \frac{2Q_L}{\sqrt{Q_{e2}Q_{e3}}}  \\
		\frac{2Q_L}{\sqrt{Q_{e1}Q_{e3}}} &  \frac{2Q_L}{\sqrt{Q_{e2}Q_{e3}}} &  1-\frac{2Q_L}{Q_{e3}} \\
		\end{array}\right)
\end{equation}
    assuming the resonant frequency of the cavity has been tuned to the axion frequency.

The output amplitude $b_1$ can be determined with the cavity $S$ matrix
\begin{equation}
\left(\begin{array}{c}
b_1\\
b_2\\
b_3\\
\end{array}\right)
=
S_{\mathrm{cavity}}\left(\begin{array}{c}
a_1\\
a_2\\
a_3\\
\end{array}\right),
\end{equation}
the effect of the amplifier and phase shifter
\begin{equation}
a_2=\sqrt{G}b_1e^{i\delta},
\end{equation}
and $a_1=0$.   I will take $\delta$ to have been chosen to be an integral multiple of $2\pi$.  This gives
\begin{equation}
b_1=\frac{2MQ_L}{\sqrt{Q_{e1}Q_{e3}}} a_3
\end{equation}
where I define $M$ to be the apparent multiplier to the $Q$ of the system
\begin{equation}
M=\left(1-2Q_L\sqrt{\frac{G}{Q_{e1}Q_{e2}}}\right)^{-1}.
\end{equation}
Note that for a passive resonator $M=1$ and that systems with $M<0$ will oscillate regardless of the presence of an axion signal. 

I note that from the signal power derived in Refs.~\cite{Cavity_idea_2,PhysRevLett.80.2043},
\begin{equation}
\frac{|a_3^2|}{Q_{e3}}=\frac{1}{4}g_{a\gamma\gamma}^2VB_0^2\rho_aC\frac{1}{m_a}
\end{equation}
where $V$ is the volume of the microwave cavity, $B_0$ is the magnetic field strength, $\rho_a$ is the density of dark matter axions, $C$ is a mode dependent form factor of order 1, and $m_a$ is the axion mass.  
$g_{a\gamma\gamma}$ is the axion-photon coupling strength, defined as $g_{a\gamma\gamma}=\frac{\alpha g_\gamma}{\pi f_a}$ where $\alpha$ is the fine structure constant, $f_a$ is the axion decay constant and $g_\gamma$ is an order unity model dependent factor.

 The power measured at the output of the system is thus
\begin{equation}
\label{eqn:rawsig}
P_{\mathrm{signal}}=\frac{M^2Q_L^2}{Q_{e1}}g_{a\gamma\gamma}^2VB_0^2\rho_a\frac{1}{m_a}
\end{equation}

For a fixed $M$, this power is maximized when the first antenna is critically coupled so that $Q_{e1}=2Q_{L}$.  It is also important to note that for the isothermal halo model of dark matter axions,  the characteristic width of the axion signal is expected to correspond of a $Q$ of $10^6$ \cite{Cavity_idea}.   Experiments with a bandwidth smaller than the axion signal width sample only a subset of axions.  Thus the conservative estimate for signal power is

\begin{equation}
\label{eqn:sig}
P_{\mathrm{signal}}=\frac{M}{2}\min\left(MQ_L,10^6\right)g_{a\gamma\gamma}^2VB_0^2\rho_a\frac{1}{m_a},
\end{equation}

though it could be larger in models where the axion dark matter is unusually cold.

The thermal noise from the physical temperature of the cavity and electronic noise of the amplifier are also magnified by the active feedback.
The noise in an active feedback resonator, $T_n$ is minimized when
\begin{equation}
\frac{1}{Q_{e1}}=\frac{G}{Q_{e2}}
\end{equation} 
and has been calculated and measured in Ref.~\cite{noise_temp_afr} to be
\begin{equation}
T_n=\frac{MQ_L}{Q_{0}}T_{\mathrm{cav}}+\frac{G}{G-1}\frac{MQ_L}{Q_{0}}T_{\mathrm{amp}}
\end{equation}
where $T_{\mathrm{cav}}$ is the physical temperature of the cavity and $T_{\mathrm{amp}}$ is the noise temperature of the amplifier.   For a system with a gain $G\gg1$, a critically coupled antenna, and the conditions $Q_{e2},Q_{e3}\ll Q_{e1}$, this becomes
\begin{equation}
T_n\approx\frac{M}{2}\left(T_{\mathrm{cav}}+T_{\mathrm{amp}}\right).
\end{equation}

The noise bandwidth will be chosen to be the axion bandwidth for experiments with $MQ<10^6$, while experiments with higher $Q$ will use their full bandwidth.  This makes the noise power

\begin{equation}
\label{eqn:noise}
P_{\mathrm{noise}}=k_B\frac{M}{2}\left(T_{\mathrm{cav}}+T_{\mathrm{amp}}\right)\frac{f_0}{\max\left(MQ_L,10^6\right)}
\end{equation}

where $k_B$ is the Boltzmann constant and $f_0$ is the resonant frequency of the cavity.

Combining equations \ref{eqn:sig} and \ref{eqn:noise} the signal to noise of an active resonator microwave cavity experiment will be
\begin{equation}
\mathrm{SNR}=M Q_L\frac{g_{a\gamma\gamma}VB_0^2\rho_aC}{k_B f_0 m_a\left(T_{\mathrm{cav}}+T_{\mathrm{amp}}\right)}
\end{equation}

Thus, despite the magnification of the noise level, the SNR of the system is proportional to $MQ_L$.  The sensitivity and scan speed of a microwave cavity dark matter axion experiment will scale with $M$.

Values of $M$ have been reported as high as $10^4$ yielding $Q$s of $7\times10^7$ in active feedback resonators~\cite{:/content/aip/journal/rsi/84/8/10.1063/1.4817537} used for other purposes, but that were operated at frequencies that could be relevant to dark matter axion searches.
This would suggest existing experiments could increase their SNR by several orders of magnitude by implementing active feedback.

The ADMX experiment is within an order of magnitude of the sensitivity necessary to be sensitive to pessimistically coupled axion dark matter.  The addition of active feedback, along with other planned upgrades, should allow it to to be sensitive to pessimistically coupled axion dark matter even in models where axions constitute a very small fraction of the dark matter.  The use of active feedback will also allow future experiments to operate at higher frequencies, where previous microwave cavity experiments have not had the sensitivity necessary to detect axion dark matter with a reasonable scan speed.

This work was supported in part by the U.S. Department of Energy under contract DE-SC0009800.


\bibliography{2014_ouroboros}

\end{document}